# Purification and characterization of an arene *cis*-dihydrodiol dehydrogenase endowed with broad substrate specificity toward PAH dihydrodiols

Jouanneau, Yves, and Meyer, Christine.

Département de Réponse et Dynamique Cellulaires/BBSI and CNRS UMR 5092 . CEA-Grenoble, F-38054 Grenoble Cedex 9, France.

Running title : A broad specificity arene dihydrodiol dehydrogenase

Corresponding author:

Yves Jouanneau

CEA-Grenoble, DRDC/BBSI,

F-38054 Grenoble Cedex 9, France.

Tel. : 33 (0)4.38 78.43.10

Fax : 33 (0)4.38 78.51.85

Email : yves.jouanneau@cea.fr

**Summary**

Initial reactions involved in the bacterial degradation of polycyclic aromatic hydrocarbons (PAHs) include a ring-dihydroxylation catalyzed by a dioxygenase, and a subsequent oxidation of the dihydrodiol products by a dehydrogenase. In this study, the dihydrodiol dehydrogenase (PDDH) from the PAH-degrading strain *Sphingomonas* CHY-1, has been characterized. The *bphB* gene encoding this enzyme was cloned and overexpressed as a His-tagged protein. The recombinant protein was purified as a homotetramer with an apparent $M_r$ of 110,000. PDDH oxidized the *cis*-dihydrodiols derived from biphenyl and eight polycyclic hydrocarbons including chrysene, benz[a]anthracene and benzo[a]pyene, to corresponding catechols. Remarkably, the enzyme oxidized pyrene 4,5-dihydrodiol, whereas pyrene is not metabolized by strain CHY-1. The PAH catechols produced by PDDH rapidly auto-oxidized in air, but were regenerated upon reaction of the *o*-quinones formed with NADH. Kinetic analyses performed under anoxic conditions revealed that the enzyme efficiently utilized two- to four-ring dihydrodiols, with $K_m$ in the range 1.4 to 7.1 $\mu$M, and exhibited a much higher Michaelis constant for $NAD^+$ ($K_m$ = 160 $\mu$M). At pH 7.0, the specificity constant ranged from 1.3 ± 0.1 $10^6$ $M^{-1}$ $s^{-1}$ with benz[a]anthracene 1,2-dihydrodiol to 20.0 ± 0.8 $10^6$ $M^{-1}$ $s^{-1}$ with naphthalene 1,2-dihydrodiol. The catalytic activity of the enzyme was 13-fold higher at pH 9.5. PDDH was subjected to inhibition by NADH and by 3,4-dihydroxyphenanthrene, and the inhibition patterns suggested that the mechanism of the reaction was ordered Bi Bi. The regulation of PDDH activity appears as a means to prevent the accumulation of PAH catechols in bacterial cells.

## Introduction

Bacteria able to grow on non-phenolic aromatic hydrocarbons generally initiate degradation by dihydroxylation of the aromatic ring, which is catalyzed by a dioxygenase (9). The reaction produces *cis*-dihydrodiols that are further oxidized to catechol derivates by a NAD-dependent dehydrogenase. The next step in the catabolic pathway is catalyzed by a second dioxygenase, which cleaves the aromatic ring (10).

Dihydrodiol dehydrogenases have been described in bacteria degrading monoaromatic compounds (1, 25, 26, 31), biphenyl and polychlorinated biphenyls (18, 32), and naphthalene (2, 23). Except for the *cis*-toluene dihydrodiol dehydrogenase from a *Bacillus* strain, which appeared to be hexameric (31), all these enzymes were found to be tetrameric and exhibited common features such as an absolute requirement for $NAD^+$. The resolution of the crystal structure of a biphenyl dihydrodiol dehydrogenase established that such enzymes belong to a subgroup of the so-called short-chain alcohol dehydrogenase/reductase (SDR) family (11). Although these enzymes are relatively well characterized from a biochemical point of view, kinetic data are rather scarce and large differences were observed upon comparison of the published Michaelis constants (see reference (31)). For example, the apparent $K_m$ for $NAD^+$ ranged from 43.5 $\mu$M for benzene dihydrodiol dehydrogenase (1) to 800 $\mu$M for naphthalene dihydrodiol dehydrogenase (23). The reported $K_m$ values for dihydrodiol substrates also vary widely between < 2 $\mu$M and higher than 200 $\mu$M. In the multi-step catabolic pathway of aromatics, the availability of kinetic data on individual enzymatic reactions is essential for a better understanding of the whole process, and for identifying possible metabolic bottlenecks. In the case of PAH degradation, little attention has been paid to the second reaction step, and no enzyme able to dehydrogenate dihydrodiols composed of more than three rings has been described so far.

In a recent study, we have identified the genes encoding a ring-hydroxylating dioxygenase from *Sphingomonas* sp. CHY-1, a strain isolated for its ability to grow on chrysene as sole carbon source (6, 36). This dioxygenase, called PhnI, was found to catalyze the dihydroxylation of various PAHs ranging in size from 2 to 5 fused rings ((6) ; Jouanneau and Meyer, unpublished results). In the present report, a gene encoding a dihydrodiol dehydrogenase has been cloned from strain CHY-1 and overexpressed in *Escherichia coli*. The biochemical properties of this enzyme, as well as its catalytic activity toward PAH dihydrodiols, are presented. The enzyme showed an exceptionally broad substrate specificity, allowing the determination, for the first time, of the kinetic parameters toward dihydrodiols derived from 2-5 ring PAHs. In addition, the mechanism of the reaction has been investigated on the basis of patterns of product inhibition.

# MATERIALS AND METHODS

## Chemicals

$NAD^+$, NADH, PAHs, antibiotics, and most other chemicals were purchased from Sigma-Aldrich (Saint-Quentin-Fallavier, France). Oligonucleotides, as well as isopropyl-β-D-thiogalactopyranoside (IPTG) were purchased from Eurogentec (Seraing, Belgium). The *cis*-dihydrodiols used in this study (Fig. 1) were prepared from cultures of *E. coli* recombinant strains overproducing the PhnI ring-hydroxylating dioxygenase from *Sphingomona*s CHY-1 (6), and incubated with a PAH or biphenyl. The purification and characterization of the diol compounds will be described elsewhere. Briefly, recombinant cells of BL21(pSD9)(pEB431) were grown on 0.8 L of LB medium and subjected to overnight induction with 0.2 mM IPTG at 25°C. Cells were harvested, resuspended in the same volume of M9 medium containing 10 mM glucose, then incubated for 24 or 48 h with 160 ml of silicone oil (Rhodorsil 47V20, Sodipro, Echirolles; France) containing 0.1g/L of PAH. The *cis*-dihydrodiols were recovered from the centrifuged culture medium by solid-phase extraction on C18 Varian Bond Elut-cartridges, and eluted in a small volume of acetonitrile. The diol compounds were further purified by reverse-phase HPLC and stored in acetonitrile in sealed vials at -20°C. A strain overproducing the pyrene dioxygenase from *Mycobacterium* strain 6PY1 (Pdo1; (21)) was used to prepare pyrene 4,5-dihydrodiol. The dihydrodiol concentrations were calculated using the following absorption coefficients: $\varepsilon_{303}$ = 13,600 $M^{-1}.cm^{-1}$ for *cis*-2,3-dihydroxy 2,3-dihydrobiphenyl (8); $\varepsilon_{262}$ = 8,114 $M^{-1}.cm^{-1}$ for *cis*-1,2-dihydroxy 1,2-dihydronaphthalene (13); $\varepsilon_{252}$ = 38,300 $M^{-1}.cm^{-1}$ and $\varepsilon_{260}$ = 43,000 $M^{-1}.cm^{-1}$ for *cis*-3,4-dihydroxy 3,4-dihydrophenanthrene (14); $\varepsilon_{244}$ = 55,600 $M^{-1}.cm^{-1}$ and $\varepsilon_{287}$ = 17,000 $M^{-1}.cm^{-1}$ for *cis*-1,2-dihydroxy 1,2-dihydroanthracene (14); $\varepsilon_{278}$ = 57,650 $M^{-1}.cm^{-1}$ for *cis*-3,4-dihydroxy 3,4-dihydrochrysene (4); $\varepsilon_{280}$ = 66,500 $M^{-1}.cm^{-1}$ for *cis*-9,10-dihydroxy-9,10-dihydrobenzo[a]pyrene (7); $\varepsilon_{263}$ = 31,000 $M^{-1}.cm^{-1}$ for *cis*-1,2- dihydroxy-1,2-

dihydrobenz[a]anthracene (15). Pyrene 4,5-dihydrodiol was estimated using an absorption coefficient of 98,0000 $M^{-1}.cm^{-1}$ at 260 nm, and fluoranthene 2,3-dihydrodiol was quantified using $\varepsilon_{263}$ = 40,000 $M^{-1}.cm^{-1}$.

**Cloning strategy and sequence analysis of *bphB***

DNA manipulations were carried out as previously described (6). Two oligonucleotides were designed after internal sequences of the *bphB* gene from *Novosphingobium aromaticivorans* (Accession number AF079317). The forward (5'-ATGTGGCATCGAAGCACGC) and reverse primer (5'-CGCGAGGCTAGCAAGGCA) corresponded to the regions 470-488 and 705-722 of the *bphB* gene sequence, respectively. PCR amplification using these primers and cosmid pSD1G3 (6) as template generated a fragment of the expected size (250-bp), suggesting that the *bphB* gene of strain CHY-1 was carried on pSD1G3. Each primer was used to determine the sequence of a 1460-bp DNA piece of cosmid pSD1G3, which was found to include the 801-bp *bphB* gene sequence (DNA sequence determination was performed by Genome express, Meylan, France). Using pSD1G3 as template, the *bphB* coding sequence was amplified with primers 5'-GGATCCAAGGAACG*CATATG*ACTGCAAG and 5'-GGATCCTCGAGCTTCAGCTTCCGGG, which introduced a NdeI site (shown in italics) and BamH1 sites (underlined) at the ends of the amplicon. The introduction of an NdeI site also allowed the replacement of the GTG start codon of *bphB* by an ATG. The PCR fragment was cloned into pGEM-T easy (Promega France, Charbonnières), giving pGHB1, and the DNA insert was checked by nucleotide sequencing. The insert showed one nucleotide change in codon 251(ATT instead of GTT) compared to the original sequence, resulting in a predicted substitution of Val for Ile. This conservative substitution was expected to have no significant influence on the properties of the gene product. The *bphB* gene was subcloned as an NdeI-BamHI fragment into pET15b (Novagen) resulting in plasmid pEBB1, which was used for overexpression of *bphB* in *E. coli* BL21(DE3). DNA sequence analyses as well as protein sequence comparison and alignment were performed using BLAST or programs available on the Infobiogen server (http://www.infobiogen.fr/deambulum/index.php).

**Overexpression of *bphB* and purification procedure**

Strain BL21(DE3)(pEBB1) was grown on 0.66 L of terrific broth at 37°C up to a bacterial density of 1.8 at 600 nm. Induction was triggered with 0.5 mM IPTG, and cells were incubated at 30°C for three hours before being harvested by centrifugation at 10,000 g for 10 min. The pellet was washed in 40 ml of 25 mM Tris-HCl, pH 8.0, and stored at –20°C. Purification was carried out at 4°C in buffers containing 5% glycerol. The cell pellet was resuspended in 10 ml of 25 mM Tris-HCl, pH 7.5, containing 0.5 M NaCl (buffer A) and treated with 0.2 mg/ml of lysozyme for 10 min at 30°C. After cooling on ice, the suspension was subjected to ultrasonication for 3 min at 30% of maximum amplitude using a Vibra cell™ ultrasonifier (Fisher Bioblock scientific, Illkirch, France). The lysate was centrifuged for 10 min at 15,000 × *g* to remove cell debris, then applied onto a 4-ml column of $Co^{2+}$-IMAC resin (TALON, BD Biosciences, Ozyme, France). The column was washed with 40 ml of buffer A, then with 16 ml of buffer A containing 20 mM imidazole. The His-tagged protein was eluted in a small volume of 25 mM Tris-HCl, pH 7.5, containing 300 mM imidazole. Active fractions were pooled, then immediately dialyzed against 800 ml of buffer A lacking NaCl. The purified protein was concentrated to 8.3 mg/ml by ultrafiltration through a 30-kDa cut-off Ultrafree centrifugal device (Millipore, Amilabo, Chassieu, France), and stored as pellets in liquid nitrogen.

**Enzyme assays**

The dehydrogenase was assayed by spectrophotometric measurement of the reduction of $NAD^+$ at 340 nm. Routine assays were performed at 30°C in a 1-ml quartz cuvette maintained under argon. The assay mixture (0.6 ml) contained 1 mM $NAD^+$, 0.1 mM of dihydrodiol substrate, in 0.1 M potassium phosphate, pH 7.0. The reaction was initiated by an appropriate amount of enzyme and the absorption at 340 nm was recorded at 0.1-s intervals over 1 min with a HP8452 spectrophotometer, equipped with a thermostated cuvette holder (Agilent Technologies, Les Ulis, France). The enzyme activity was calculated from the initial linear portion of the time course using an absorption coefficient of 6,220 $M^{-1}.cm^{-1}$ for NADH. One enzyme unit was defined as the amount that catalyzed the formation of one micromole of NADH per min. When appropriate, the absorption coefficient was corrected to take into account the contribution of the PAH catechol products at 340 nm, using the following values ($M^{-1}.cm^{-1}$): 1,650 for I,2-dihydroxynaphthalene; 2,770 for 3,4-dihydroxyphenanthrene; 2,750 for 1,2-dihydroxyanthracene; 3,850 for 3,4-dihydroxychrysene. When 2,3-dihydroxy-2,3-dihydrobiphenyl was used as a substrate, $NAD^+$ reduction was recorded at 360 nm ($\varepsilon_{360(}$ = 4,570 $M^{-1}.cm^{-1}$), because this diol absorbed at 340 nm. The strong absorbance of the oxidation products from benz[a]anthracene 1,2-dihydrodiol and benzo[a]pyrene 9,10-dihydrodiol in the 300-380 nm range precluded accurate determination of $NAD^+$ reduction. Instead, the enzyme activity was assayed by measuring the formation of the oxidation product at 316 nm in the former case, at 398 nm in the latter. At 316 nm, the absorption coefficients of the 1,2-dihydroxybenz[a]anthracene, the diol and NADH were calculated to be 10,800, 1,240 and 4,160 $M^{-1}.cm^{-1}$, respectively. Some assays were performed at pH 9.5 under similar experimental conditions except that phosphate buffer was replaced by 50 mM bicine. For determination of the pH optimum, pH was varied by 0.5 increments with phosphate as buffer in the range 7.0-8.0, and with bicine in the range 8.0-10.0. *cis*-2,3-dihydroxy-2,3-dihydrobiphenyl was the substrate, and the enzyme concentration was 11.5 nM.

**Kinetic experiments**

For each dihydrodiol, the steady-state kinetic parameters of the PDDH-catalyzed reaction were determined from sets of enzyme assays where the substrate concentration was varied in a 1-100 $\mu$M range. All assays were run in duplicates and at least 8 concentrations were tested per substrate. The enzyme concentration was adjusted between 17.5 and 70 nM depending on substrate. For reactions performed at variable concentrations of $NAD^+$, the dihydrodiol concentration was kept constant at 20 $\mu$M, and $NAD^+$ was varied in a 10-1200 $\mu$M range. the The Michaelis-Menten equation was fitted to the plots of the initial reaction rate versus substrate concentration using the curve fit option of KaleidaGraph (Synergy Software). Only curve fits showing correlation coefficients better than 0.98 were considered.

**Chemical identification of the enzymatic products**

To obtain sufficient product for chemical analysis, the dehydrogenase catalyzed reaction was carried out in a 3-ml closed cuvette containing up to 150 $\mu$M of substrate in a final volume of 2 ml. The reaction was initiated by an appropriate amount of enzyme, and allowed to reach completion, at which time ethyl acetate extraction was carried out twice with an equal volume of solvent. After drying with sodium sulfate and solvent evaporation under nitrogen, the dry material was dissolved in acetonitrile or tetrahydrofuran. Samples were derivatized either with bis(trimethysilyl)trifluoroacetamide : trimethylchlorosilane (99:1) or with *n*-butylboronate (NBB), both reagents from Supelco (Sigma-Aldrich), as previously described (21), then subjected to GC-MS analysis using a HP6890/HP5973 apparatus (Agilent Technologies). Samples (2.5 $\mu$l) were injected in the split mode (a split ratio of 20:1 or 10:1 was used) on a capillary column (30-m long, 0.25-mm internal diameter, Varian VF-5ms), which was developed with a temperature gradient from 75 to 300°C, as previously described (21). Mass spectrum acquisitions were carried out in the total ion current mode.

**Protein analyses**

Routine protein assays were performed with the bicinchoninic acid reagent kit (Pierce) using bovine serum albumin as a standard. SDS-PAGE on mini-slab gels was performed as previously described (16). The molecular mass of the purified dehydrogenase was determined by chromatography on a HR 10/30 column of Superdex SD200 (Amersham Biosciences). The column was run at a flow rate of 0.2 ml/min and calibrated with the following protein markers, all from Sigma-Aldrich : Ferritin (443 kDa), catalase (240 kDa), aldolase (150 kDa), bovine serum albumin (67 kDa), ovalbumin (43 kDa) and myoglobin (17 kDa).

**Nucleotide sequence accession number:** The nucleotide sequence described in this report has been deposited in the DDBJ/EMBL/Genbank data bases under accession number AM230449.

## RESULTS

**Cloning and sequence analysis of the *bphB* gene from strain CHY-1**

In a previous study, we have identified two clusters of catabolic genes possibly involved in PAH biodegradation in strain CHY-1 (6). These genes were similar in sequence and arrangement to counterparts found on a megaplasmid in *Novosphingobium aromaticivorans* strain F199 (27). The megaplasmid of stain F199 was found to carry 79 catabolic genes including a *bphB* gene encoding a putative dihydrodiol dehydrogenase. Using primers designed after two conserved sequences of *bphB*, a fragment of the expected size (250 bp) was amplified by PCR with genomic DNA from CHY-1 as template. An identical fragment was generated when cosmid pSD1G3 was used as template, suggesting that the gene of interest was borne on the 40-kb cosmid insert. Nucleotide sequence analysis of a 1460-bp DNA piece of pSD1G3 revealed the presence of a *bphB* gene flanked by *xylA*-like and *xylC*-like genes (partial sequences, data not shown). This arrangement was similar to that found in strain F199. The *bphB* gene of CHY-1 consists of 801-bp, features a GTG as start codon, and encodes a polypeptide having a predicted molecular mass of 27,731 Da. This polypeptide exhibited highest sequence similarities with corresponding gene products found in sphingomonads, including *N. aromaticivorans* F199 (91% identity; accession number: AF079317) and *Sphingomonas* BN6 (76%; U65001). It also showed significant similarity with dehydrogenases genetically linked to PAH catabolism in various genera including *Burkholderia* (PhnB, 51% identity; AF061751), *Nocardioides* (PhdE, 51%; AB031319), *Ralstonia* (NagB, 48%; AF036940), *Pseudomonas* (NahB, 47%; AF125184), and *Rhodococcus* (NarB, 48%; AY392424). A lower degree of similarity was observed with the BphB gene products found in PCB-degrading bacteria like *Rhodococcus* RHA1 (48% identity; D32142), *R. globerulus* P6 (43%; P47230), *Comamonas testosteroni* B-356 (44%; Q46381), and *Burkholderia xenovorans* LB400 (43%; (11)).

**Purification and properties of the recombinant His-tagged dihydrodiol dehydrogenase**

The *bphB* coding sequence was subcloned in pET15b resulting in a gene fusion encoding a protein product bearing an N-terminal His-tag. Overexpression of this fusion in *E. coli* BL21(DE3) gave rise to the accumulation in the cells of a protein showing a subunit size of 30-kDa upon SDS-PAGE analysis. The His-tagged enzyme was purified by IMAC chromatography, resulting in an apparently homogenous preparation as judged from SDS-PAGE. The purification procedure yielded approximately 5 mg of enzyme per g of bacteria. The molecular mass of the protein, called PDDH, was determined to be 103 kDa by gel filtration chromatography, suggesting that it is a tetramer. The recombinant enzyme showed high dehydrogenase activity with 2,3-dihydroxy 2,3-dihydrobiphenyl as substrate and $NAD^+$ as cofactor. Its activity was 25-fold lower when $NADP^+$ (1 mM) was used as cofactor. PDDH exhibited a pH optimum close to 9.5. Although the activity at this pH was 13-fold higher than at pH 7.0, we carried out most enzymatic assays at pH 7.0 because it is closer to the physiological conditions. In many respects, the dihydrodiol dehydrogenase from strain CHY-1 resembled counterparts involved in aromatic hydrocarbon degradation in other bacteria, especially biphenyl and naphthalene degraders (23, 34).

**Substrate reactivity of PDDH**

Since the PhnI ring-hydroxylating dioxygenase from strain CHY-1 is able to convert a wide range of PAHs to the corresponding dihydrodiols, it was of interest to examine whether PDDH would further oxidize these arene diols (Fig. 1). For this purpose, the dihydrodiols prepared by incubating cells overproducing PhnI with 2- to 5-ring PAHs, were provided as substrates to the dehydrogenase. As summarized in tables 1 and 2, all the arene dihydrodiols produced by PhnI were substrates of PDDH. Benz[a]anthracene was oxidized by PhnI to give three dihydrodiols, one of which was identified as the 1,2-isomer based on comparison of UV-absorbance and GC-MS spectra with previously published data (15). The other two diols were tentatively identified as the 8,9- and the 10,11-isomers based on the same criteria (data not shown). Interestingly, all three isomers were converted to the corresponding catechols by PDDH. Furthermore, PDDH was also able to oxidize pyrene 4,5-dihydrodiol although PhnI was unable to attack pyrene (6). On the other hand, no enzyme activity was observed with either phenanthrene 9,10-dihydrodiol or chrysene 3,4,9,10-*bis*-*cis*-dihydrodiol. The latter compound was recently identified as an oxidation product of chrysene generated by PhnI (Jouanneau and Meyer, unpublished results).

**Identification of the dehydrogenase reaction products**

The conversion of arene *cis*-dihydrodiols to catechols catalyzed by PDDH was first revealed by UV-absorbance changes as exemplified by the spectral changes of benzo[a]pyrene dihydrodiol upon oxidation by the dehydrogenase (Fig. 2B). Due to the strong absorbance of this particular dihydrodiol at 340 nm, which precluded accurate measurement of NADH, the enzyme activity was determined by measuring the formation of the catecholic product at 398 nm ($\varepsilon_{398}$ = 14,500 $M^{-1}.cm^{-1}$). With benz[a]anthracene 1,2-dihydrodiol as substrate, the activity was determined from absorbance recordings at 316 nm, which corresponded to a minimum in the spectrum of this diol (Fig. 2A). The activity was calculated from the sum of the contributions of NADH and of the catecholic product, corrected for the dihydrodiol absorption at this wavelength (see Materials and methods).

The PAH catechols formed by PDDH were identified by GC-MS analysis of trimethylsilyl derivates (Table 1). The spectra all showed a dominant signal at m/z values corresponding to the expected molecular ions, and fragmentation patterns consistent with catecholic structures. N-butylboronate derivatisation yielded products whose mass spectra confirmed the catecholic nature of the compounds formed by PDDH (data not shown). All catechols derived from PAH dihydrodiols were found to be unstable, and underwent auto-oxidation to *o*-quinones, as illustrated in the case of the 3,4-dihydroxyphenanthrene (Fig. 3A). Based on absorbance changes observed during oxidation, the half-lives of 3,4-dihydroxyphenanthrene, 3,4-dihydroxychrysene, 1,2-dihydroxynaphthalene and 1,2-dihydroxyanthracene in air-saturated buffer at pH 7.0 were estimated to be 20 min, 30 min, 40 s and < 30 s, respectively. Besides, the *o*-quinones formed readily reacted with NADH to regenerate the catechols, as shown in the case of the phenanthrene 3,4-dione (Fig 3B). The instability of the PAH catechols, which justified that PDDH assays be carried out under anoxic conditions, complicated the analysis of these compounds by GC-MS. Indeed, preliminary attempts to identify the catecholic derivates of naphthalene and anthracene were unsuccessful, most likely because these products auto-oxidize during the extraction or the derivatization procedure. We found that the catechols could be stabilized in tetrahydrofuran, and that the stabilizing effect was due to the antioxidant additive present in commercial THF, namely 2,6-di-*tert*-butyl *p*-cresol. At a concentration of 0.025% in the solvent, this agent prevented the auto-oxidation of all catechols tested and even promoted the reverse reaction, that is the conversion of *o*-quinone to catechol. Once stabilized, 1,2-dihydroxynaphthalene and 1,2-dihydroxyanthracene could be identified by GC-MS (Table 1).

**Steady-state kinetics**

Under the given assay conditions, PDDH exhibited Michaelis-Menten behaviour when the concentration of the studied *cis*-arene dihydrodiols was varied in the range 1-100 $\mu$M. The steady-state parameters calculated for six diol substrates are presented in Table 2. The apparent $K_m$ increased slightly when the number of fused rings of the substrate increased, but the values were remarkably close for substrates with quite different structures. Likewise, the enzyme reached almost the same maximum velocity regardless of the substrate size for 2-3 ring dihydrodiols and the 4-ring *cis*-3,4-dihydroxy 3,4-dihydrochrysene. However, the dihydrodiols derived from benz[a]anthracene, pyrene, fluoranthene, and benzo[a]pyrene were oxidized at slower rates (Table 2). In the latter three cases, only the maximum initial rates of substrate conversion are reported, as steady state kinetics were not determined. At the pH optimum of the enzyme, the maximum velocity increased 13-fold and 11-fold with biphenyl dihydrodiol and chrysene dihydrodiol as substrates, respectively. Comparison of the specificity constants indicated that naphthalene dihydrodiol was the best substrate, whereas

benz[a]anthracene 1,2-dihydrodiol appeared to be the least preferred, essentially because of a higher apparent $K_m$ of the enzyme for this substrate.

The $K_m$ of PDDH for $NAD^+$ was determined from steady-state kinetic experiments where the nucleotide concentration was varied in the range 0.01-1.2 mM at a saturating concentration of dihydrodiol. The $K_m$ value was estimated to be 160 ± 16 $\mu$M at pH 7.0 using biphenyl dihydrodiol (0.05 mM) as substrate, and 168 ± 16 $\mu$M with phenanthrene 3,4-dihydrodiol (0.02 mM) as substrate.

**Product inhibition**

The inhibitory effects of NADH and 3,4-dihydroxyphenanthrene on the catalytic activity of PDDH were studied. The dihydroxyphenanthrene was freshly prepared from the complete conversion of the diol with PDDH and subsequent extraction with ethyl acetate. The concentration of the product, kept in tetrahydrofuran, was estimated using an absorption coefficient of 23,800 $M^{-1}.cm^{-1}$ at 280 nm. When the concentration of phenanthrene dihydrodiol was varied and that of NADH was kept constant, in the presence of increasing amounts of dihydroxyphenanthrene, a mixed type inhibition was observed in that the catecholic product caused both a decrease in the $V_{max}$ and an increase in the $K_m$ (Fig. 4). The calculated $K_m/V_{max}$ ratios were found to be related to the inhibitor concentration through a parabolic function. When the concentration of $NAD^+$ was varied at a constant and nearly saturating concentration of dihydrodiol (25 $\mu$M), the inhibition exerted by dihydroxyphenanthrene on PDDH activity was essentially non-competitive (Fig. 5). When the $K_m/V_{max}$ ratio was plotted against the inhibitor concentration, a straight line was obtained, and the intersect with the abscissa gave an inhibition constant of 10.9 $\mu$M. These results indicated that dihydroxyphenanthene acted as both a noncompetitive and a dead-end inhibitor. Finally, NADH was found to be a competitive inhibitor with respect to $NAD^+$, and an inhibition constant of 41.5 $\mu$M was determined from a plot of the $K_m/V_{max}$ ratios versus the NADH concentration. The product inhibition patterns described above are consistent with an enzyme having an ordered Bi-Bi mechanism, in which $NAD^+$ binds first and NADH is released last. Similarly ordered Bi Bi mechanisms have been reported for many other alcohol dehydrogenases (5). However, the mechanism of the PDDH enzyme is not ordinary in that the inhibition by the catecholic product is of mixed type instead of being merely noncompetitive, a pattern which has been previously observed for the benzyl alcohol dehydrogenase from *Pseudomonas putida* (29).

## DISCUSSION

The enzyme described in this study appears as one of the most versatile *cis*-dihydrodiol dehydrogenase ever reported. It is the first example of an enzyme able to oxidize dihydrodiols derived from two- to five-ring PAHs, including the carcinogenic benz[a]anthracene and benzo[a]pyrene. In comparison, the naphthalene dihydrodiol dehydrogenase from *P. putida* was found to oxidize substrates with 2- and 3-rings, but it failed to transform substrates with four or five rings (23). On the other hand, the BphB enzyme from PCB degraders appeared to be more efficient toward biphenyl dihydrodiol and chlorinated derivates, although the activity of that enzyme toward PAH dihydrodiols other than naphthalene dihydrodiol has not been reported (2).

In *Sphingomonas* strain CHY-1, PDDH is associated with a ring-hydroxylating dioxygenase endowed with extended catalytic capabilities toward PAHs (6). It is shown here that all PAH dihydrodiols produced by PhnI in strain CHY-1 are readily converted to catechols by PDDH. Remarkably, all three isomers of benz[a]anthracene dihydrodiol generated by PhnI are substrates of PDDH. The substrate range of the latter enzyme appeared to exceed that of the former, since pyrene 4,5-dihydrodiol was oxidized by PDDH whereas PhnI was inactive on pyrene. The conversion of pyrene dihydrodiol to the corresponding *o*-quinone was previously observed by *S. yanoikuyae* R1, suggesting that a dehydrogenase similar to PDDH might occur in this strain (17). On the other hand, diol compounds formed as byproducts of non-productive dioxygenase reactions, including phenanthrene 9.10-dihydrodiol generated by certain dioxygenases (e.g. a pyrene dioxygenase described in (21)), and chrysene *bis-cis* dihydrodiol formed by PhnI, were not substrates of PDDH.

The Michaelis constants presented in Table 2 suggested that PDDH had a high affinity for most of the PAH dihydrodiols tested, and indicated that the kinetic parameters of the enzyme were remarkably similar irrespective of the substrate size. Comparison of the specificity constants indicated that naphthalene 1,2-dihydrodiol was the best substrate for the enzyme, as found for other dehydrogenases in *Pseudomonas* strains degrading naphthalene (2, 23). However, the PDDH constant for this substrate was higher by one or two orders of magnitude than that previously reported, essentially because of differences in the $K_m$ values (2, 23). Remarkably, the specificity constant of PDDH for biphenyl dihydrodiol was as high as that of dehydrogenases found in strains known for their ability to degrade biphenyl and polychlorobiphenyls (2, 34). The enzyme velocity of PDDH at its pH optimum (321 $s^{-1}$) was comparable to or higher than that previously published for related enzymes (2, 30, 34).

The catalytic activity of PDDH, and especially its broad specificity, is determined by its tridimensional structure, which certainly resembles that of the *B. xenovorans* LB400 enzyme, since the two proteins share 43% sequence identity. The active site of the LB400 enzyme features a catalytic triad, composed of residues Ser142, Tyr155 and Lys159, which is conserved among enzymes of the SDR family (11). The equivalent residues in PDDH are Ser144, Tyr157 and Lys161. The critical role of these residues in catalysis was demonstrated experimentally by Vedadi *et al.* using variants of the *C. testosteroni* enzyme (34). The Asn143 residue of the LB400 enzyme has been predicted to be a determinant of substrate specificity, based on a model of the enzyme-substrate complex where this residue was found to bind one hydroxyl group of the substrate (11). This prediction also comes from the observation that the BphB enzymes from PCB degraders all contain an asparagine residue at this position, whereas in enzymes from naphthalene degraders, this residue was replaced by a valine (11). Consistent with this idea, the NahB dehydrogenase from *P. putida* G7, which belongs to the latter category, showed a 30-fold higher $K_m$ for biphenyl dihydrodiol than BphB from *C. testosteroni* (2). The counterpart of Asn143 in the PDDH enzyme is Thr145, the side chain of which might potentially form a H-bond with the substrate. In contrast to the naphthalene dihydrodiol dehydrogenases discussed above, the $K_m$ of PDDH for biphenyl dihydrodiol is similar to that of BphB enzymes. While the role of this threonine residue in substrate binding remains to be assessed experimentally, a data base search for sequences homologous to PDDH in Sphingomonads showed that all but one contained a threonine in identical position.

From a physiological point of view, the kinetic properties of PDDH reported herein have implications with regards to the metabolism of PAHs by strain CHY-1 and related *Sphingomonas* strains (6, 24). Given that the initial dioxygenase PhnI has a low turnover (>2 $s^{-1}$; Jouanneau and Meyer, unpublished results) similar to biphenyl dioxygenase (12), and that PDDH showed low $K_m$ values for PAH dihydrodiols, it is likely that the dihydrodiols are rapidly converted to catechols. The bacteria then have to cope with the formation of highly unstable PAH catechols, which rapidly auto-oxidize to *o*-quinones. As shown in Fig.3, this auto-oxidation might be prevented, to some extent, by NADH. However, catechols produced *in vivo* by dehydrogenation of PAH dihydrodiols have been shown to induce oxidative DNA damage to the cells (28). In addition, single-electron redox reactions with quinones could generate reactive oxygen species that are deleterious to cells (3). Hence, bacteria able to metabolize PAHs must have evolved strategies to neutralize the potentially toxic effect of catechols and *o*-quinones. Detoxification systems relying on the activity of catechol-*o*-methyltransferases and *o*-quinone reductases were found in PAH-degrading *Mycobacterium* strains (19, 20). While it is unknown whether similar enzymes are present in Sphingomonads, glutathione S-transferases have been proposed to have a detoxification function in relation to PAH catabolism, because genes encoding such enzymes were frequently encountered associated to catabolic genes involved in aromatic and polycyclic hydrocarbon degradation in these bacteria (22, 35).

The PDDH activity was strongly inhibited by NADH as indicated by the observed low $K_i$ value (42 $\mu$M). It was also inhibited by the catecholic product of the reaction, which likely acts as an uncompetitive and a dead-end inhibitor, as exemplified by the effect of dihydroxyphenanthrene (Fig. 4). Both types of regulation might concur in controlling the formation of catechols, and in preventing their accumulation inside the cells during PAH catabolism. A buildup of the PAH catechol concentration would occur as a consequence of a metabolic block at the third step of the PAH catabolic pathway. This step is catalyzed by an extradiol dioxygenase, a generally labile enzyme which may undergo oxidative inactivation during the catalytic processing of certain substrates (33). Hence, the metabolic regulation of PDDH might be essential to adjust enzyme activity to physiological needs and prevent cell intoxication. A similar mechanism of inhibition has been described for toluene dihydrodiol dehydrogenase (31), and benzyl alcohol dehydrogenase (29), suggesting that such a regulation is widespread among bacteria degrading aromatic hydrocarbons.

**Acknowledgments**

We thank John Willison for helpful discussions and critical reading of the manuscript. This work was supported by grants from the Centre National de la Recherche Scientifique, and the Commisariat à l'Energie Atomique to UMR5092.

## References

1. **Axcell, B. C., and P. J. Geary.** 1973. The metabolism of benzene by bacteria. Purification and some properties of the enzyme *cis*-1,2-dihydroxycyclohexa-3,5-diene (nicotinamide adenine dinucleotide) oxidoreductase (*cis*-benzene glycol dehydrogenase). Biochem. J. **136:**927-934.

2. **Barriault, D., M. Vedadi, J. Powlowski, and M. Sylvestre.** 1999. *cis*-2,3-dihydro-2,3-dihydroxybiphenyl dehydrogenase and *cis*-1,2-dihydro-1,2-dihydroxynaphathalene dehydrogenase catalyze dehydrogenation of the same range of substrates. Biochem. Biophys. Res. Commun. **260:**181-187.

3. **Bolton, J. L., M. A. Trush, T. M. Penning, G. Dryhurst, and T. J. Monks.** 2000. Role of quinones in toxicology. Chem. Res. Toxicol. **13:**135-160.

4. **Boyd, D. R., N. D. Sharma, R. Agarwal, S. M. Resnick, M. J. Schocken, D. T. Gibson, J. M. Sayer, H. Yagi, and D. M. Jerina.** 1997. Bacterial dioxygenase-catalysed dihydroxylation and chemical resolution routes to enantiopure *cis*-dihydrodiols of chrysene. J. Chem. Soc. Perkin Trans. 1**:**1715-1723.

5. **Brändén, C.-I., H. Jörnvall, H. Eklund, and B. Furugren.** 1975. Alcohol dehydrogenases, p. 103-190. *In* P. D. Boyer (ed.), The enzymes, 3rd ed, vol. 11. Academic Press, New York.

6. **Demaneche, S., C. Meyer, J. Micoud, M. Louwagie, J. C. Willison, and Y. Jouanneau.** 2004. Identification and functional analysis of two aromatic ring-hydroxylating dioxygenases from a *Sphingomonas* strain degrading various polycyclic aromatic hydrocarbons. Appl. Environ. Microbiol. **70:**6714-6725.

7. **Gibson, D. T., V. Mahadevan, D. M. Jerina, H. Yogi, and H. J. Yeh.** 1975. Oxidation of the carcinogens benzo [a] pyrene and benzo [a] anthracene to dihydrodiols by a bacterium. Science **189:**295-297.

8. **Gibson, D. T., R. L. Roberts, M. C. Wells, and V. M. Kobal.** 1973. Oxidation of biphenyl by a *Beijerinckia* species. Biochem. Biophys. Res. Commun. **50:**211-219.

9. **Gibson, D. T., and V. Subramanian.** 1984. Microbial degradation of aromatic hydrocarbons, p. 181-252. *In* D. T. Gibson (ed.), Microbial degradation of organic compounds. Dekker, M., New York.

10. **Harayama, S., M. Kok, and E. L. Neidle.** 1992. Functional and evolutionary relationships among diverse oxygenases. Annu. Rev. Microbiol. **46:**565-601.

11. **Hülsmeyer, M., H. J. Hecht, K. Niefind, B. Hofer, L. D. Eltis, K. N. Timmis, and D. Schomburg.** 1998. Crystal structure of *cis*-biphenyl-2,3-dihydrodiol-2,3-dehydrogenase from a PCB degrader at 2.0 A resolution. Protein Sci. **7:**1286-1293.

12. **Imbeault, N. Y., J. B. Powlowski, C. L. Colbert, J. T. Bolin, and L. D. Eltis.** 2000. Steady-state kinetic characterization and crystallization of a polychlorinated biphenyl-transforming dioxygenase. J. Biol. Chem. **275:**12430-12437.

13. **Jeffrey, A. M., H. J. Yeh, D. M. Jerina, T. R. Patel, J. F. Davey, and D. T. Gibson.** 1975. Initial reactions in the oxidation of naphthalene by *Pseudomonas putida*. Biochemistry **14:**575-584.

14. **Jerina, D. M., H. Selander, H. Yagi, M. C. Wells, J. F. Davey, V. Mahadevan, and D. T. Gibson.** 1976. Dihydrodiols from anthracene and phenanthrene. J. Am. Chem. Soc. **98:**5988-5996.

15. **Jerina, D. M., P. J. Vanbladeren, H. Yagi, D. T. Gibson, V. Mahadevan, A. S. Neese, M. Koreeda, N. D. Sharma, and D. R. Boyd.** 1984. Synthesis and absolute-configuration of the bacterial *cis*-1,2-dihydrodiol, *cis*-8,9-dihydrodiol, and *cis*-10,11-dihydrodiol metabolites of benz[a]anthracene formed by a strain of *Beijerinckia*. J. Org. Chem. **49:**3621-3628.

16. **Jouanneau, Y., C. Meyer, I. Naud, and W. Klipp.** 1995. Characterization of an *fdxN* mutant of *Rhodobacter capsulatus* indicates that ferredoxin I serves as electron donor to nitrogenase. Biochim. Biophys. Acta **1232:**33-42.

17. **Kazunga, C., and M. D. Aitken.** 2000. Products from the incomplete metabolism of pyrene by polycyclic aromatic hydrocarbon-degrading bacteria. Appl. Environ. Microbiol. **66:**1917-1922.

18. **Khan, A. A., R. F. Wang, M. S. Nawaz, and C. E. Cerniglia.** 1997. Nucleotide sequence of the gene encoding *cis*-biphenyl dihydrodiol dehydrogenase (*bphB*) and the expression of an active recombinant His-tagged *bphB* gene product from a PCB degrading bacterium, *Pseudomonas putida* OU83. FEMS Microbiol. Lett. **154:**317-324.

19. **Kim, Y. H., K. H. Engesser, and C. E. Cerniglia.** 2003. Two polycyclic aromatic hydrocarbon *o*-quinone reductases from a pyrene-degrading *Mycobacterium*. Arch. Biochem. Biophys. **416:**209-217.

20. **Kim, Y. H., J. D. Moody, J. P. Freeman, B. Brezna, K. H. Engesser, and C. E. Cerniglia.** 2004. Evidence for the existence of PAH-quinone reductase and catechol-O-methyltransferase in *Mycobacterium vanbaalenii* PYR-1. J. Ind. Microbiol. Biotechnol. **31:**507-516.

21. **Krivobok, S., S. Kuony, C. Meyer, M. Louwagie, J. C. Willison, and Y. Jouanneau.** 2003. Identification of pyrene-induced proteins in *Mycobacterium* sp. 6PY1 : Evidence for two ring-hydroxylating dioxygenases. J. Bacteriol. **185:**3828-3841.

22. **Lloyd-Jones, G., and P. C. Lau.** 1997. Glutathione S-transferase-encoding gene as a potential probe for environmental bacterial isolates capable of degrading polycyclic aromatic hydrocarbons. Appl. Environ. Microbiol. **63:**3286-3290.

23. **Patel, T. R., and D. T. Gibson.** 1974. Purification and propeties of (+)-*cis*-naphthalene dihydrodiol dehydrogenase of *Pseudomonas putida*. J. Bacteriol. **119:**879-888.

24. **Pinyakong, O., H. Habe, and T. Omori.** 2003. The unique aromatic catabolic genes in sphingomonads degrading polycyclic aromatic hydrocarbons (PAHs). J. Gen. Appl. Microbiol. **49:**1-19.

25. **Raschke, H., T. Fleischmann, J. R. Van der Meer, and H. P. E. Kohler.** 1999. *cis*-Chlorobenzene dihydrodiol dehydrogenase (TcbB) from *Pseudomonas* sp strain P51, expressed in *Escherichia coli* DH5 alpha(pTCB149), catalyzes enantioselective dehydrogenase reactions. Appl. Environ. Microbiol. **65:**5242-5246.

26. **Reiner, A. M.** 1972. Metabolism of aromatic compounds in bacteria. Purification and properties of the catechol-forming enzyme, 3,5-cyclohexadiene-1,2-diol-1-carboxylic acid (NAD + ) oxidoreductase (decarboxylating). J. Biol. Chem. **247:**4960-4965.

27. **Romine, M. F., L. C. Stillwell, K. K. Wong, S. J. Thurston, E. C. Sisk, C. Sensen, T. Gaasterland, J. K. Fredrickson, and J. D. Saffer.** 1999. Complete sequence of a 184-kilobase catabolic plasmid from *Sphingomonas aromaticivorans* F199. J. Bacteriol. **181:**1585-1602.

28. **Seike, K., M. Murata, K. Hirakawa, Y. Deyashiki, and S. Kawanishi.** 2004. Oxidative DNA damage induced by benz[a]anthracene dihydrodiols in the presence of dihydrodiol dehydrogenase. Chem. Res. Toxicol. **17:**1445-1451.

29. **Shaw, J., M. Rekik, F. Schwager, and S. Harayama.** 1993. Kinetic studies on benzyl alcohol dehydrogenase encoded by TOL plasmid pWWO. A member of the zinc-containing long chain alcohol dehydrogenase family. J. Biol. Chem. **268:**10842-10850.

30. **Shaw, J. P., and S. Harayama.** 1990. Purification and characterisation of TOL plasmid-encoded benzyl alcohol dehydrogenase and benzaldehyde dehydrogenase of *Pseudomonas putida*. Eur. J. Biochem. **191:**705-714.

31. **Simpson, H. D., J. Green, and H. Dalton.** 1987. Purification and some properties of a novel heat-stable *cis*-toluene dihydrodiol dehydrogenase. Biochem. J. **244:**585-590.

32. **Sylvestre, M., Y. Hurtubise, D. Barriault, J. Bergeron, and D. Ahmad.** 1996. Characterization of active recombinant 2,3-dihydro-2,3-dihydroxybiphenyl dehydrogenase from *Comamonas testosteroni* B-356 and sequence of the encoding gene (*bphB*). Appl. Environ. Microbiol. **62:**2710-2715.

33. **Vaillancourt, F. H., G. Labbe, N. M. Drouin, P. D. Fortin, and L. D. Eltis.** 2002. The mechanism-based inactivation of 2,3-dihydroxybiphenyl 1,2-dioxygenase by catecholic substrates. J. Biol. Chem. **277:**2019-2027.

34. **Vedadi, M., D. Barriault, M. Sylvestre, and J. Powlowski.** 2000. Active site residues of *cis*-2,3-dihydro-2,3-dihydroxybiphenyl dehydrogenase from *Comamonas testosteroni* strain B-356. Biochemistry **39:**5028-5034.

35. **Vuilleumier, S.** 1997. Bacterial glutathione S-transferases: what are they good for ? J. Bacteriol. **179:**1431-1441.

36. **Willison, J. C.** 2004. Isolation and characterization of a novel sphingomonad capable of growth with chrysene as sole carbon and energy source. FEMS Microbiol. Lett. **241:**143-150.

## Figure legends

Figure 1 : Chemical structures of the dihydrodiols that were used as substrates by the PDDH enzyme from *Sphingomonas* CHY-1. PDDH also utilized benz[a]anthracene dihydrodiol isomers bearing hydroxyls in positions 8,9 and 10,11, but failed to transform 9,10-dihydroxy 9,10-dihydrophenanthrene..

Figure 2: Spectral changes observed upon dehydrogenation of benz[a]anthracene 1,2-dihydrodiol (A) benzo[a]pyrene dihydrodiol (B) catalyzed by PDDH. The reactions were carried out under argon in a closed cuvette containing a suboptimal concentration of $NAD^+$ in 1.0 ml of reaction mixture, pH 7.0. This allowed absorbance recording down to 260 nm.
In panel A, the concentration of $NAD^+$ and benz[a]anthracene 1,2-dihydrodiol were made 0.25 mM and 40 $\mu$M, respectively. The reaction was started at time zero by injecting 14 nM of PDDH. In panel B, the cuvette contained 0.1 mM $NAD^+$, and 8 $\mu$M of benzo[a]pyrene dihydrodiol. The reaction was initiated at time zero by 35 nM of PDDH. Spectra were recorded at the times indicated (seconds), after enzyme addition. The initial and final spectra are shown as a thicker line.

Figure 3: Auto-oxidation of 3,4-dihydroxyphenanthrene in air-saturated buffer and reduction of the *o*-quinone by NADH. The dihydroxyphenanthrene was prepared under anoxic conditions by incubating 0.15 mM of 3,4-dihydrodiol in the presence of 0.25 mM $NAD^+$ and 41.5 $\mu$g of PDDH for 20 min in 2 ml of reaction buffer. It was extracted with ethyl acetate under argon, then dissolved in 0.1 ml of acetonitrile. A sample of this solution (10 $\mu$l; 30 nmol) was immediately diluted in 1 ml of air-saturated 0.1 M phosphate buffer, pH 7.0, in a cuvette. Panel A : Spectra were recorded at the following time points : 0, 1, 5, 10, 30, 60, 120 min. The spectrum of dihydroxyphenanthrene is shown as a thicker line

Panel B : The cuvette was degassed for 15 min under argon, and the *o*-phenanthrene quinone (shown as a thick line) was reduced by successive NADH additions, 17 nmol each, at time 0, 10, and 25 min. Spectra were recorded at the time intervals indicated (min). After 45 min, the product was extracted as above. Its absorbance spectrum was found to be identical to that of 3,4-dihydroxyphenanthrene (data not shown).

Figure 4: Inhibition of PDDH by 3,4-dihydroxyphenanthrene at a constant concentration of $NAD^+$. Enzyme assays were carried out in the absence (●) or in the presence of 3,4-dihydroxyphenanthrene at the concentrations of 7.5 (■), 15 (♦) or 25 $\mu$M (▲). $NAD^+$ concentration was kept constant at 0.15 mM. Solid lines represent fits of the data to the Michaelis equation obtained using the curve fit function of Kaleidagraph.

Figure 5 : Inhibition of PDDH by 3,4-dihydroxyphenanthrene at a constant concentration of phenanthrene dihydrodiol. Enzyme assays were carried out in the absence (●) or in the presence of 3,4-dihydroxyphenanthrene at the concentrations of 2.5 (■),), 5 (◆), 10 (▲), or 20 $\mu$M (▼). The dihydrodiol concentration was kept constant at 25 $\mu$M. Data were fitted by the Michaelis equation as in Figure 4. A plot of the calculated $K_m/V_{max}$ ratios as a function of inhibitor concentration gave a straight line (not shown), and an inhibition constant of 10.9 $\mu$M was calculated.

TABLE 1: GC-MS identification of the PAH catechols produced by PDDH

| | | Properties of the TMS-derivates of the products |
|---|---|---|
| Substrate | RT (min) | Fragment ions (% relative intensity, assignment) |
| Biphenyl 2,3-dihydrodiol | 15.66 | 330(100, $M^+$), 315(20.5, $M^+$-$CH_3$), 257(8.5, $M^+$-TMS), 242(47.2, $M^+$-TMS-$CH_3$), 227(41.8, $M^+$-TMS- 2$CH_3$), 212(48.6, $M^+$-TMS- 3$CH_3$), 152(10.0, $M^+$-2OTMS) |
| Naphthalene 1,2-dihydrodiol | 13.81 | 304(100, $M^+$), 289(4.2, $M^+$-$CH_3$), 216(40.7, $M^+$-TMS-$CH_3$), 201(5.6, $M^+$-TMS-2$CH_3$), 186(25.7, $M^+$-TMS- 3$CH_3$) |
| Phenanthrene 3,4-dihydrodiol | 19.47 | 354(100, $M^+$), 339(3.1, $M^+$-$CH_3$), 281(3.0, $M^+$-TMS), 266(54.7, $M^+$-TMS-$CH_3$), 236(50.5, $M^+$-TMS-3$CH_3$), 176(5.1, $M^+$-2OTMS) |
| Anthracene 1,2-dihydrodiol | 20.54 | 354(100, $M^+$), 339(2.5, $M^+$-$CH_3$), 281(2.2, $M^+$-TMS), 266(58.0, $M^+$-TMS-$CH_3$), 236(33.5, $M^+$-TMS-3$CH_3$), 176(3.2, $M^+$-2OTMS) |
| Pyrene 4,5-dihydrodiol | 22.46 | 378(100, $M^+$), 363(4.4, $M^+$-$CH_3$), 291(17.8), 290(51.7, $M^+$-TMS-$CH_3$), 260(32, $M^+$-TMS-3$CH_3$), 215(11.0), 207(13.8), 200(3.2, $M^+$-2OTMS) |
| Fluoranthene 2,3-dihydrodiol | 22.14 | 378(100, $M^+$), 363(3.6, $M^+$-$CH_3$), 291(8.2), 290(35.6, $M^+$-TMS-$CH_3$), 260(23.7, $M^+$-TMS-3$CH_3$), 215(6.1), 207(5.1), 200(2.9, $M^+$-2OTMS) |
| Chrysene 3,4-dihydrodiol | 25.04 | 404(100, $M^+$), 389(3.9, $M^+$-$CH_3$), 331(4.8, $M^+$-TMS), 317(16.6), 316(70.0, $M^+$-TMS-CH3), 286(45.0, $M^+$-TMS-3$CH_3$), 226(6.9, $M^+$-2OTMS) |
| Benz[a]anthracene 1,2-dihydrodiol | 23.82 | 404(100, $M^+$), 389(2.6, $M^+$-$CH_3$), 331(1.5, $M^+$-TMS), 316(33.1, $M^+$-TMS-$CH_3$), 286(19.7, $M^+$-TMS-3$CH_3$), 226(2.2, $M^+$-2OTMS) |
| Benzo[a]pyrene dihydrodiol | 28.06 | 428(100, $M^+$), 355(3.8, $M^+$-TMS), 340(33.8, $M^+$-TMS-$CH_3$), 310(14.6, $M^+$-TMS-3$CH_3$), 282(9.1, $M^+$-2TMS), 281(14.0), 250(3.9, $M^+$-2OTMS) |

TABLE 2: Apparent steady-state kinetic parameters of PDDH for dihydrodiols

| Substrate | $K_m$ | $V_{max}$ | $k_{cat}$ | $k_{cat}/K_m$ |
|---|---|---|---|---|
| | $\mu$M | U.mg$^{-1}$ | s$^{-1}$ | $\times 10^6$ M$^{-1}$ s$^{-1}$ |
| Naphthalene 1,2-dihydrodiol | 1.41± 0.25 | 14.2 ± 0.5 | 28.2 ± 1.0 | 20.0 ± 0.8 |
| Phenanthrene 3,4-dihydrodiol | 1.85 ± 0.45 | 8.4 ± 0.5 | 16.7 ± 1.0 | 9.0 ± 0.7 |
| Anthracene 1,2-dihydrodiol | 2.00 ± 0.35 | 10.9 ± 1.0 | 21.6 ± 1.9 | 10.8 ± 1.1 |
| Biphenyl 2,3-dihydrodiol | | | | |
| pH 7.0 | 3.50 ± 0.45 | 12.4 ± 0.30 | 24.6 ± 0.6 | 7.0 ± 0.2 |
| pH 9.5 | 3.60 ± 0.6 | 162 ± 6 | 321 ± 12 | 89.2 ± 4.0 |
| Chrysene 3,4-dihydrodiol | | | | |
| pH 7.0 | 2.20 ± 0.25 | 10.2 ± 0.35 | 20.3 ± 0.7 | 9.2 ± 0.35 |
| pH 9.5 | 5.70 ± 1.6 | 110 ± 12 | 218 ± 24 | 38.2 ± 5.9 |
| Benz[a]anthracene 1,2-dihydrodiol | 7.1 ± 1.3 | 4.57 ± 0.35 | 9.1 ± 0.7 | 1.3 ± 0.1 |
| Pyrene 4,5-dihydrodiol[a] | | 3.2 | 6.35 | |
| Fluoranthene 2,3-dihydrodiol[a] | | 2.65 | 5.25 | |

| | | |
|---|---|---|
| Benzo[a]pyrnene dihydrodiol[a] | 4.0 | 7.9 |

[a] The values reported were calculated from the maximum initial rates of either NADH formation (pyrene and fluoranthene diols as substrates), or product formation (benzo[a]pyrene diol).

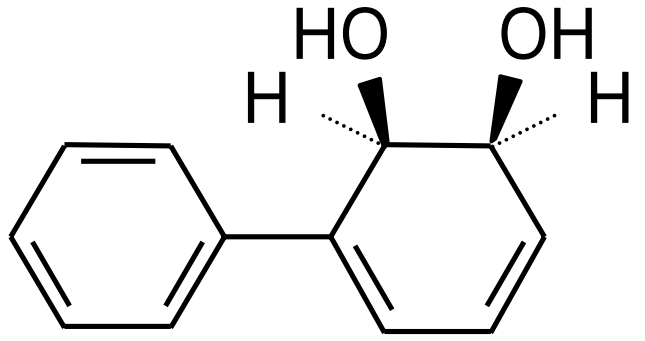

2,3-dihydroxy
2,3-dihydrobiphenyl

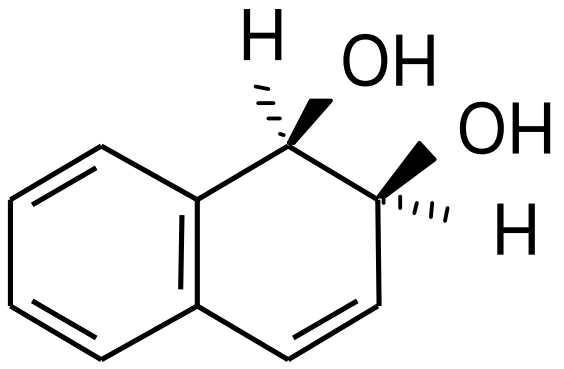

1,2-dihydroxy
1,2-dihydronaphthalene

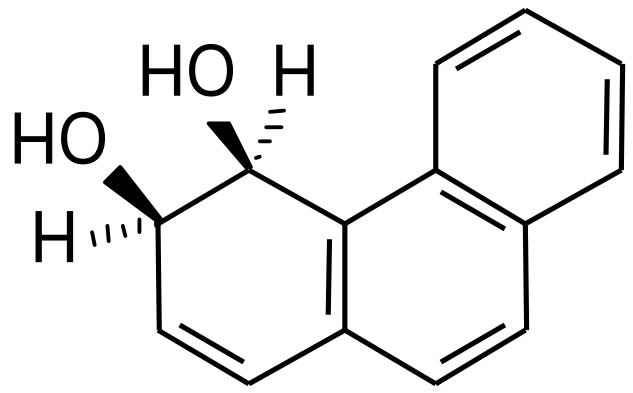

3,4-dihydroxy
3,4-dihydrophenanthrene

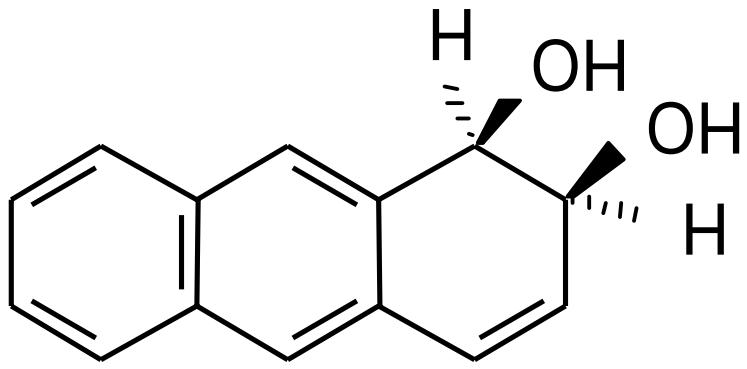

1,2-dihydroxy
1,2-dihydroanthracene

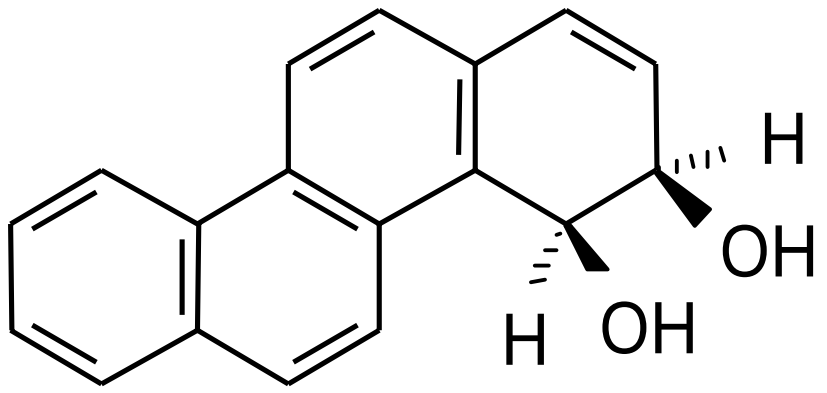

3,4-dihydroxy 3,4-dihydrochrysene

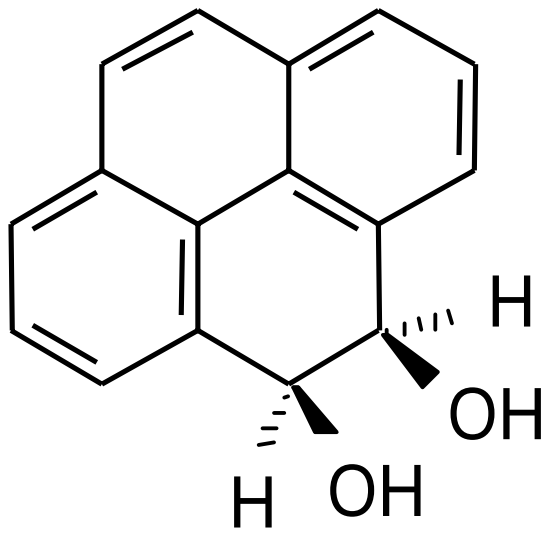

4,5-dihydroxy 4,5-dihydropyrene

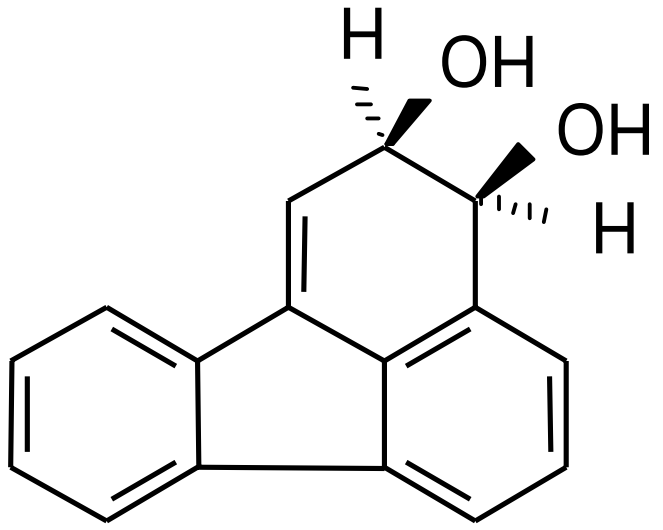

2,3-dihydroxy
2,3-dihydrofluoranthene

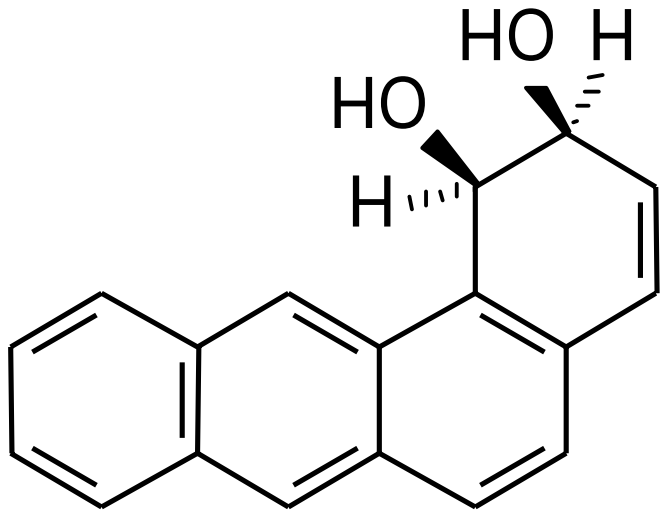

1,2-dihydroxy
1,2-dihydrobenz[a]anthracene

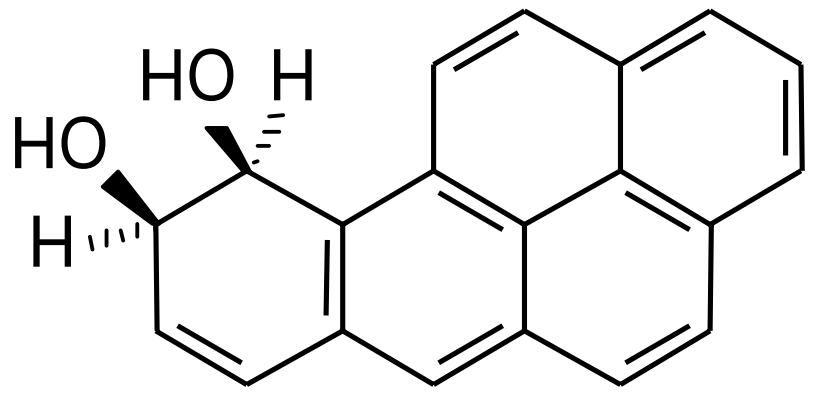

9,10-dihydroxy
9,10-dihydrobenzo[a]pyrene

Fig.1

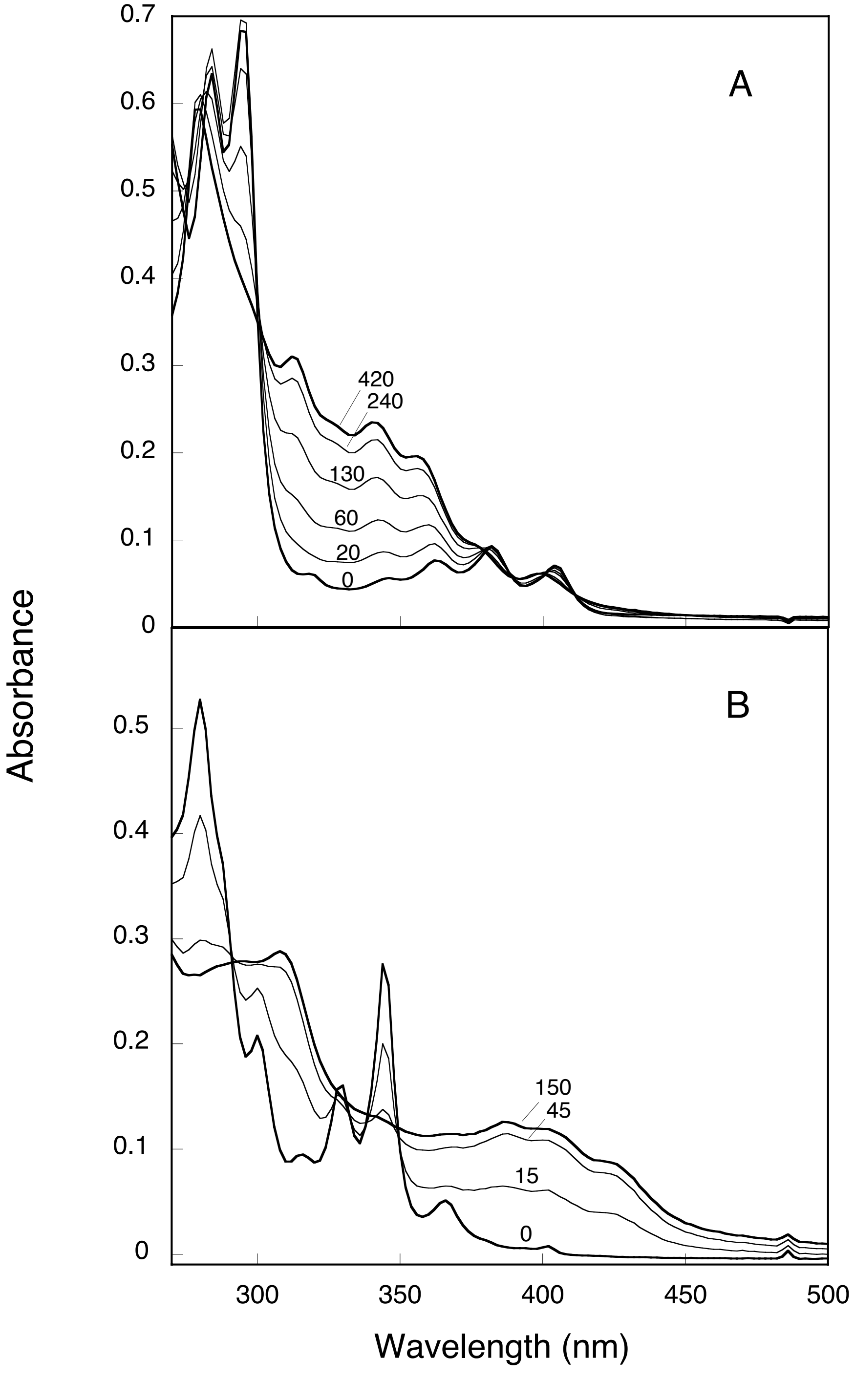

Figure 2

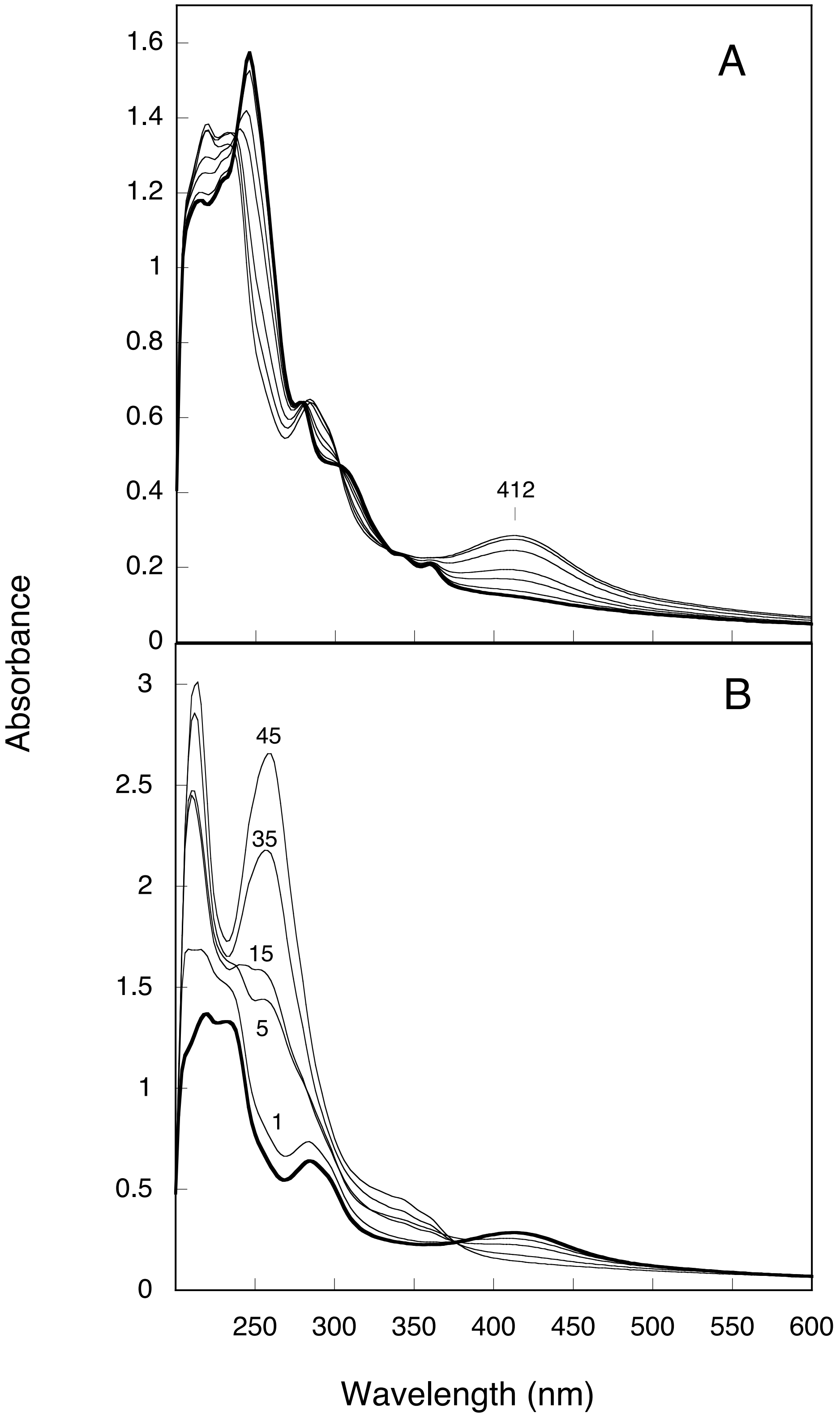

Figure 3

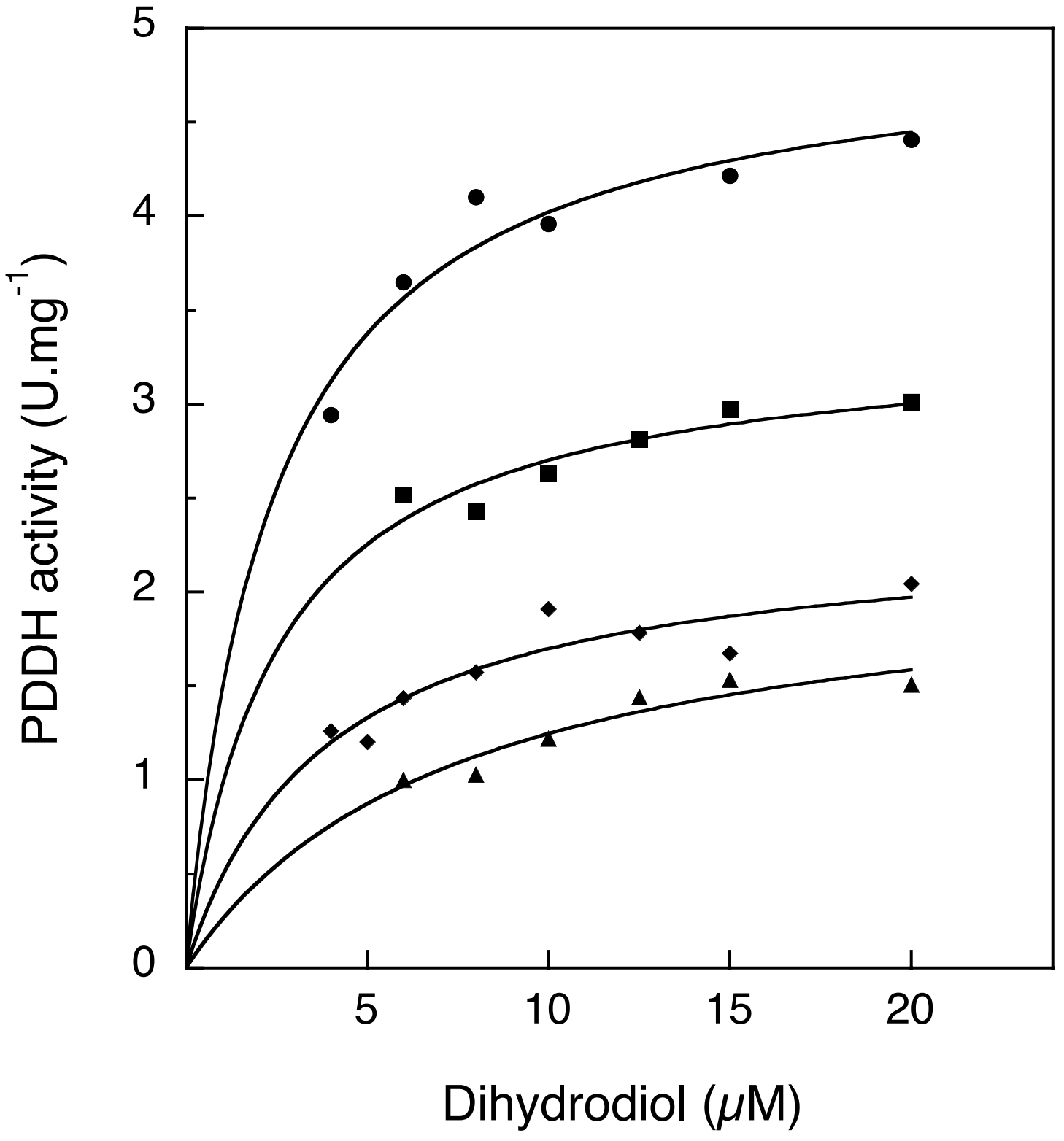

Fig. 4

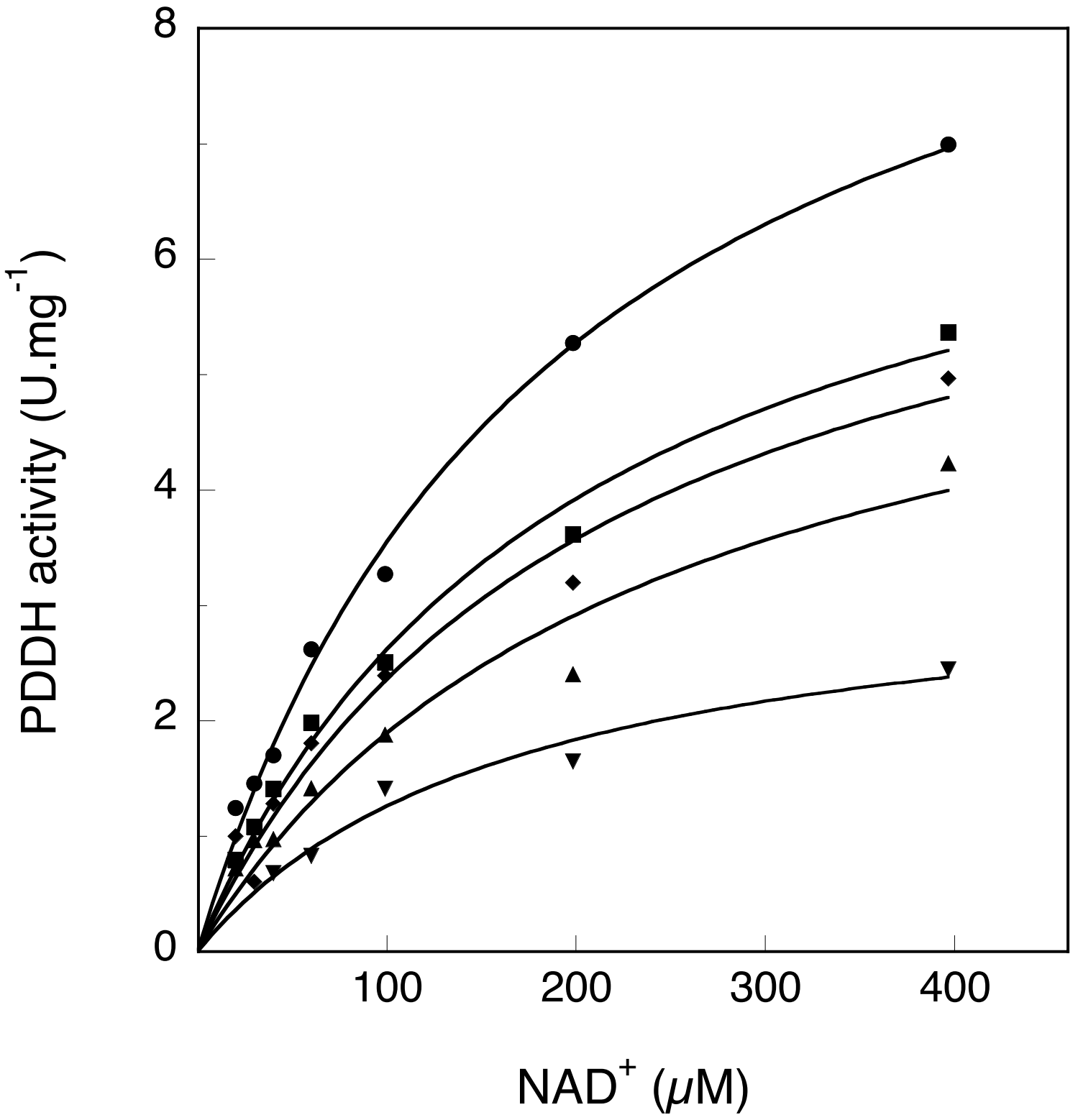

Fig. 5